\documentclass[a4paper, titlepage, oneside, 11pt]{amsart}
\usepackage{amsmath,amssymb,amsfonts,amstext,amscd,array,cite,enumitem,geometry,graphicx,latexsym,mathptmx,xcolor,pstricks}
\usepackage[utf8]{inputenc}
\usepackage{newunicodechar}
\usepackage[all,cmtip]{xy}

\newtheorem{lem}{Lemma}[section]

\newtheorem{thm}[lem]{Theorem}
\newtheorem{cor}[lem]{Corollary}

\newtheorem{df}[lem]{Definition}
\newtheorem{rem}[lem]{Remark}

\setlist[description]{leftmargin=\parindent,labelindent=0cm}

\newcommand{\la}{\lambda}
\def\R{{\mathbb{R}}}

\def\a{{\alpha}}
\def\b{{\beta}}

\newcommand{\vertiii}[1]{{\left\vert\kern-0.25ex\left\vert\kern-0.25ex\left\vert #1
    \right\vert\kern-0.25ex\right\vert\kern-0.25ex\right\vert}}

\numberwithin{equation}{section}

\allowdisplaybreaks

\begin{document}

\title{Yang-Baxter maps and independence preserving property}

\author[M.~Sasada]{Makiko Sasada}
\address{Graduate School of Mathematical Sciences, University of Tokyo, 3-8-1, Komaba, Meguro-ku, Tokyo, 153--8914, Japan}
\email{sasada@ms.u-tokyo.ac.jp}

\author[R.~Uozumi]{Ryosuke Uozumi}
\address{Department of Applied Physics, Graduate School of Engineering, University of Tokyo,
7-3-1 Hongo, Bunkyo-ku, Tokyo 113-8656, Japan}
\email{uozumi@fs.t.u-tokyo.ac.jp}

\begin{abstract}
We study a surprising relationship between two properties for bijective functions $F : \mathcal{X} \times \mathcal{X} \to \mathcal{X} \times \mathcal{X}$ for a set $\mathcal{X}$ which are introduced from very different backgrounds. One of the property is that $F$ is a Yang-Baxter map, namely it satisfies the \lq\lq set-theoretical" Yang-Baxter equation, and the other property is the independence preserving property (IP property for short), which means that there exist independent (non-constant) $\mathcal{X}$-valued random variables $X,Y$ such that $U,V$ are also independent with $(U,V)=F(X,Y)$. 
Recently in the study of invariant measures for a discrete integrable system, a class of functions having these two properties were found. Motivated by this,  we analyze a relationship between the Yang-Baxter maps and the IP property, which has never been studied as far as we are aware, focusing on the case $\mathcal{X}=\R_+$. Our first main result is that \textit{all} quadrirational Yang-Baxter maps $F : \R_+ \times \R_+ \to \R_+ \times \R_+$ in the most interesting subclass have the independence preserving property. In particular, we find new classes of bijections having the IP property. Our second main result is that these newly introduce bijections are fundamental in the class of (known) bijections with the IP property, in the sense that most of known bijections having the IP property are derived from these maps by taking special parameters or performing some limiting procedure. This reveals that the IP property, which has been investigated for specific functions individually, can be understood in a unified manner.
\end{abstract}



\date{\today}

\maketitle


\section{Introduction}

\subsection{Overview}

In this paper, we study a surprising relationship between two properties for bijective functions $F : \mathcal{X} \times \mathcal{X} \to \mathcal{X} \times \mathcal{X}$ for a set $\mathcal{X}$ which are introduced from very different backgrounds and seemingly unrelated. One of the property is that $F$ is a  Yang-Baxter map, namely it satisfies the \lq\lq set-theoretical" Yang-Baxter equation
\[
F_{12} \circ F_{13} \circ F_{23} = F_{23} \circ F_{13} \circ F_{12}
\]
where $F_{ij}$ acts on the $i$-th and $j$-th factors of the product $\mathcal{X} \times \mathcal{X} \times \mathcal{X}$. Yang-Baxter maps are also called set-theoretical solutions to the quantum Yang-Baxter equation. 
The other is the independence preserving property (IP property for short), which means that there is a class of quadruplets of (non-Dirac) probability distributions $\mu, \nu,\tilde{\mu}, \tilde{\nu}$ satisfying $F( \mu \times \nu) = \tilde{\mu} \times \tilde{\nu}$. In other words, there exist independent (nonconstant) $\mathcal{X}$-valued random variables $X,Y$ such that $U,V$ are also independent with $(U,V):=F(X,Y)$. The IP property has been mostly studied for explicit functions $F$ with $\mathcal{X}=\R$ or an open interval of $\R$ and used to characterize special probability distributions such as normal, gamma, exponential, inverse-Gaussian, beta and so on (more details are discussed below). Such property is also called Matsumoto-Yor property if $F$ is given by a special form. 

Recently, these characterization results with the IP property are also used to study stationary solutions of stochastic integrable models \cite{CN, CSjsp} and invariant measures of discrete integrable systems \cite{CSirf}. In this context, it was pointed out that the map
\[
F_{\mathrm{GIG}}^{\a,\b}(x,y)=\left(y \frac{1+\b xy }{1+ \a xy},  \ x \frac{1+\a xy }{1+ \b xy} \right)
\]
on $\R_+ \times \R_+$ into itself with parameters $\a,\b \ge 0$, which originates from the discrete KdV equation, have these two properties, namely they are (parameter-dependent) Yang-Baxter maps and also have the IP property for the generalized inverse Gaussian (GIG) distributions in \cite{CSirf}. This motivates us to study a relationship between the Yang-Baxter maps and the IP property, which has never been studied as far as we are aware. We note that later in \cite{BN, LW2}, it is proved that $U$ and $V$ given by $(U,V)=F_{\mathrm{GIG}}^{\a,\b}(X,Y)$ are independent only when $X$ and $Y$ have GIG distributions with proper parameters, namely the map characterizes the GIG distribution. 

Besides this specific example, recent developments on the study of probabilistic properties of discrete integrable systems lead us to expect a deep relationship between these two properties. Actually, one important consequence of the IP property is that the dynamics on the two-dimensional lattice defined by the map $F$, which typically turns out to be a discrete integrable system for such map, has a spatially independent and identically/alternately distributed invariant measure (See Theorem 2.1 of \cite{CSirf}. Note that the IP property is called the detailed balance condition there). Hence, one may expect that there is a connection between the IP property and the integrability, while the Yang-Baxter equation is one of the typical ways to characterize integrability. 

As a first study in this direction, we focus on the case $\mathcal{X}=\R_+$ in this paper, but we do not lose that much generality as discussed in Subsection \ref{subsec:setting}. There are two main contributions of our paper to connect these two properties. First, we prove that \textit{all} quadrirational Yang-Baxter maps $F : \R_+ \times \R_+ \to \R_+ \times \R_+$ in the most interesting subclass $[2:2]$ (whose definition is given in Section \ref{sec:YB-map}), which was introduced in \cite{Adler} and also studied in \cite{PSTV}, that are $F_{\mathrm{GIG}}=(F_{\mathrm{GIG}}^{\a,\b})$, $H_{I}^+=(H_{I}^{+,\a,\b}), H_{II}^+=(H_{II}^{+,\a,\b})$ and $H_{III,A}=(H_{III,A}^{\a,\b})$ given as 
\begin{align*}
H^{+}_{I} (x,y) & = \left(\frac{y}{\a}\frac{\b +\a x +\b y +\a \b xy}{1+x+y+\b xy},  \ \frac{x}{\b}\frac{\a +\a x +\b y +\a \b xy}{1+x+y+\a xy} \right) \\
H^{+}_{II} (x,y) &= \left(\frac{y}{\a}\frac{\b +\a x +\b y }{1+x+y},  \ \frac{x}{\b}\frac{\a +\a x +\b y}{1+x+y} \right) \\
H_{III,A} (x,y) &= \left(\frac{y}{\a}\frac{\a x +\b y }{x+y},  \ \frac{x}{\b}\frac{\a x +\b y }{x+y} \right)
\end{align*}
with parameters $\a,\b > 0$, have the IP property as stated in the next theorem. In \cite{PSTV}, $F_{\mathrm{GIG}}$ is named $H_{III,B}$. We give more background of these functions in Section \ref{sec:YB-map}. In the following, for any $p, q >0$, $\mathrm{Be}' (\la, a, b  ;  p, q), \mathrm{K} (\la, a, b  ;  p, q)$ and $\mathrm{GIG} (\la, a, b  ;  p, q)$ are some generalizations of Beta prime, Kummer of Type $2$ and GIG distributions respectively. The explicit density function of each probability distribution is given in Subsection \ref{subsec:notation}. The IP property for $F_{\mathrm{GIG}}=H_{III,B}$ was obtained in \cite{CSirf} so we omit it in the theorem.

\begin{thm}\label{thm:ip}
Let $\a, \b >0$. For the following distributions $(X,Y)$, the random variables $U,V$ given by $(U,V)=F(X,Y)$ for each map are independent and have the following distributions. 

(i) For $F= H^{+,\a,\b}_{I}$ : 
\begin{align*}
X \sim \mathrm{Be}'  (\la, a, b \ ; \ \a, 1), & \quad   Y \sim \mathrm{Be}'  (-\la, a, b \ ; \ \b, 1), \\
U \sim \mathrm{Be}' (-\la, a, b \ ; \ \a, 1), & \quad   V \sim \mathrm{Be}'  (\la, a, b \ ; \ \b, 1)
\end{align*} 
where $\lambda \in \R$, $a,b >0$, $- \min\{a,b\} < \frac{\lambda}{2} < \min\{a,b\}$.

(ii) For $F= H^{+,\a,\b}_{II}$ : 
\begin{align*}
X \sim \mathrm{K} (\la, a, b \ ; \ \a, 1), & \quad    Y \sim \mathrm{K}  (-\la, a, b \ ; \ \b, 1), \\
U \sim \mathrm{K} (-\la, a, b \ ; \ \a, 1), & \quad   V \sim \mathrm{K} (\la, a, b \ ; \ \b, 1)
\end{align*} 
where $\lambda \in \R$, $a,b >0$, $-b < \frac{\lambda}{2} < b$.

(iii) For $F= H^{\a,\b}_{III,A}$ : 
\begin{align*}
X \sim \mathrm{GIG} (\la, a, b \ ; \ \a, 1), & \quad    Y \sim \mathrm{GIG} (-\la, a, b \ ; \ \b, 1), \\
U \sim \mathrm{GIG}(-\la, a, b \ ; \ \a, 1), & \quad   V \sim \mathrm{GIG} (\la, a, b \ ; \ \b, 1).
\end{align*} 
where $\lambda \in \R$, $a,b >0$.
\end{thm}
By change of variables, the IP property for $H^{\a,\b}_{III,A}$ is easily reduced to the property for $F^{\a,\b}_{\mathrm{GIG}}$, but the result for $H^{+,\a,\b}_{I}$ and $H^{+,\a,\b}_{II}$ are not reduced to any known cases (except for special parameters), hence we obtain new classes having the IP property. We note that Koudou and Wesolowski also found the IP property for $H^{+,\a,\b}_{II}$ independently \cite{KW2}. We conjecture that these maps characterize each probability distribution, which was actually proved in the literature for some special cases such as $F_{\mathrm{GIG}}^{\a,\b}$ (and so true for $H^{\a,\b}_{III,A}$). The characterization for $H_{II}^{+,\a,\b}$ was proved by Koudou and Wesolowski in \cite{KW2}, which was announced just after the first version of the present paper was announced. As seen from the explicit expressions, there are beautiful similarity between these maps and the parameters of distributions, and in fact, we can prove the claims (ii) and (iii) of Theorem \ref{thm:ip} by a certain limiting procedure from the result (i), besides the direct computation. This limiting procedure is studied in Section \ref{sec:proof}. 

Our second contribution is that, as Theorem \ref{thm:relation} in Section \ref{sec:special}, we reveal that all of known functions $F$ on a product of open intervals of $\R$ having the IP property with two exceptions (which characterize normal and Cauchy distributions, both having full support on $\R$) are derived from the quadrirational Yang-Baxter maps $H^{+}_{I}, H^{+}_{II}$ or $H_{III,A}$ by taking a special parameter or performing a singular limit with an appropriate coordinate-wise change of variables. These relations are summarized in the following figure, which is the consequence of Lemma \ref{lem:scale} and Theorem \ref{thm:relation}. 
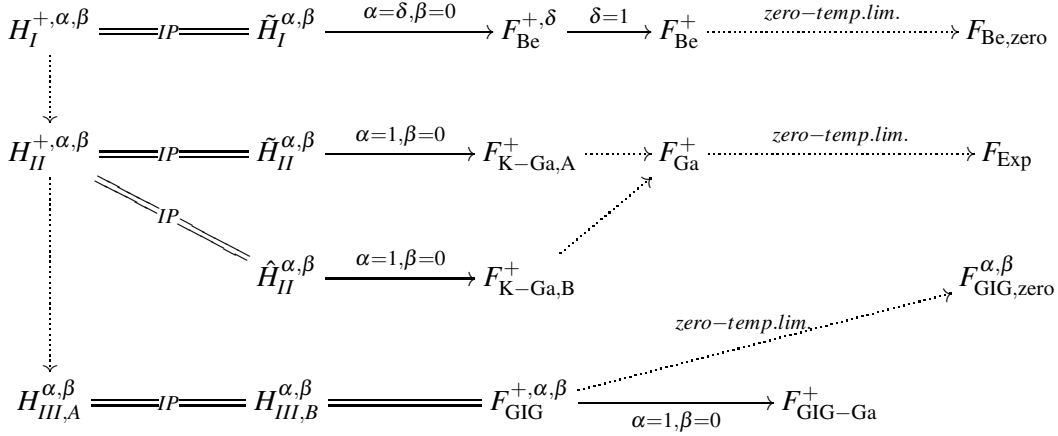
\[
   \xymatrix{
H^{+,\a,\b}_I  \ar@{=}[rr]|{IP}  \ar@{.>}[d] &&  \tilde{H}^{\a,\b}_I \ar[rr]^-{\a=\delta, \b=0} && F^{+,\delta}_{\mathrm{Be}}  \ar[r]^-{\delta=1} &  F^{+}_{\mathrm{Be}}  \ar@{.>}[rr]^-{zero-temp. lim.} & & F_{\mathrm{Be,zero}} \\ H^{+,\a,\b}_{II} \ar@{=}[rr]|{IP}  \ar@{=}[rrd]|{IP}  \ar@{.>}[dd] &&  \tilde{H}^{\a,\b}_{II} \ar[rr]^-{\a=1, \b=0} && F^{+}_{\mathrm{K-Ga,A}} \ar@{.>}[r]  &  F^+_{\mathrm{Ga}} \ar@{.>}[rr]^-{zero-temp. lim.} & & F_{\mathrm{Exp}} \\
 & & \hat{H}^{\a,\b}_{II} \ar[rr]^-{\a=1, \b=0} && F^{+}_{\mathrm{K-Ga,B}} \ar@{.>}[ru]  & && F^{\a,\b}_{\mathrm{GIG,zero}} \\ H^{\a,\b}_{III,A}  \ar@{=}[rr]|{IP}   &&  H^{\a,\b}_{III,B} \ar@{=}[rr]  & &  F^{+,\a,\b}_{\mathrm{GIG}}  \ar[rr]_-{\a=1, \b=0} \ar@{.>}[rrru]^-{zero-temp. lim.}   &&  F^+_{\mathrm{GIG-Ga}} }
\]
The dotted lines represent singular scaling limits, the equality with label $IP$ means two functions are IP-equivalent, whose definition is introduced in Section \ref{sec:IP}, and the arrows represent special parameter selections. The explicit expression of bijections in the diagram are given in Section \ref{sec:IP}. This excellently unifies most of known results for the IP property, which is summarized in Corollary \ref{cor:ip}. 

To summarize these two results in short, for the case $\mathcal{X}=\R_+$, all Yang-Baxter maps (in an interesting class) have the IP property and (most of) the functions having the IP property are obtained from a Yang-Baxter map, which was a big surprise for us.

Finally, we discuss possible future developments. First, since $H^+_{I}, H^+_{II}$ and $H_{III,A}$ are all of subtraction-free form, as already pointed out in several contexts \cite{KNW, PSTV, CSirf, CSjsp}, there is a zero-temperature version (also called a tropical version as well as an ultra-discretized version) of them where the $(+, \times)$-algebra is replaced by $(\min, +)$-algebra. For $F^{\a,\b}_{\mathrm{GIG}}$, such generalization was already studied in \cite{CSirf, BN}. These zero-temperature versions also satisfy the Yang-Baxter property as well as the IP property. However, since the IP property may also hold with some discrete distributions for such zero-temperature versions, the characterization of distributions may become more complicated. Another promising generalization is the positive definite matrix versions of $H^{+}_{I}, H^{+}_{II}$ and $H_{III,A}$. Actually, for most of bijections $F$ having the IP property discussed in Section \ref{sec:IP}, its positive definite matrix version has been introduced and shown that they also have the IP property (cf. \cite{LW, HR, KOUDOU, Koo, LW2}). The matrix version of $F^{\a,\b}_{\mathrm{GIG}}$ was introduced in \cite{LW2} and its IP property was already shown there. A very natural and interesting question is whether these matrix versions are Yang-Baxter maps. Recently, the matrix versions of stochastic integrable systems, such as random polymers and interacting diffusions are introduced \cite{OC21, ABO} and the matrix versions of the IP property (for a certain function) play an essential role. Of course, the most important and fundamental question is to figure out why there is a relationship between the Yang-Baxter maps and the IP property. Also, using the connection between the Yang-Baxter maps and the IP property, it is worth trying to find a more direct connection between quantum, stochastic and classical integrable systems and their stationary distributions in general. 

\subsection{Outline of the remaining part of the paper} In Section \ref{sec:YB-map}, we explain a brief background on the Yang-Baxter maps and review the result on the characterization of the quadrirational Yang-Baxter maps on $\mathbb{CP}^1$ as well as the origin of $H_{I}^{+}, H_{II}^{+} $ and $H_{III, A}$. In Section \ref{sec:IP}, we review known results on the IP property for bijections on a product of open intervals of $\R$ and classify them into some classes. Then, with a few exceptions of bijections, we introduce normalized versions of them by cooridnate-wise change of variables. Here, the normalized version means that it is a birational map on $\R_+ \times \R_+$ into itself. In Section \ref{sec:proof}, we prove Theorem \ref{thm:ip}, which is essentially a consequence of direct computation, and also discuss several relations between the maps $H_{I}^{+}, H_{II}^{+} $ and $H_{III,A}$ as well as the probability distributions appeared in Theorem \ref{thm:ip}. Finally, in Section \ref{sec:special}, we give explicit relations between the functions discussed in Section \ref{sec:IP}, which are already known in literature, and newly introduced maps $H_{I}^{+}, H_{II}^{+} $ and $H_{III,A}$ to derive known IP properties from Theorem \ref{thm:ip}. 

\subsection{Notation and probability distributions}\label{subsec:notation}
Here, we list frequently used bijections: 
\begin{itemize}
\item$I: \R_+ \to \R_+$, $I(x)=\frac{1}{x}$, $I^{-1}=I$,
\item For $\a>0$, ${\theta}_{\a} :\R_+ \to \R_+$, $\theta_{\a}(x)=\a x$, $(\theta_{\a})^{-1}=\theta_{\a^{-1}}$,
\item $\pi : \R_+^2 \to \R_+^2$, $\pi(x,y)=(y,x)$, $\pi^{-1}=\pi$.
\end{itemize}

We also introduce three classes of probability distributions with two positive parameters $p, q >0$. The specific parametrization is the key to connect probability measures by change of variables and scaling limits in Lemmas \ref{lem:trans}, \ref{lem:scale2}, and unify the known results on the IP property in Corollary \ref{cor:ip}. 

\begin{description}
\item[Generalized Beta prime distribution $(p, q)$]  For $\lambda, a,b \in \R$, $-b < \frac{\lambda}{2} < a$, the \emph{Generalized Beta prime} distribution with parameters $(\lambda,a,b ; p, q)$, which we denote $\mathrm{Be}'(\lambda,a,b ; p, q)$, has density
\[\frac{1}{Z} x^{\lambda-1} (1+p x)^{-a-\frac{\lambda}{2}} (1+q x^{-1})^{-b+\frac{\lambda}{2}},\qquad x\in\mathbb{R}_+,\]
where $Z$ is a normalizing constant.

\item[Kummer distribution of Type 2 $(p, q)$] For $\lambda,b \in \R$, $a >0$, $-b < \frac{\lambda}{2}$, the \emph{Kummer} distribution of Type 2 with parameters  $(\lambda,a,b ; p, q)$, which we denote $\mathrm{K} (\lambda,a,b ; p, q)$, has density
\[\frac{1}{Z} x^{\lambda-1}e^{-a p x} (1+q x^{-1})^{-b+\frac{\lambda}{2}} ,\qquad x\in\mathbb{R}_+,\]
where $Z$ is a normalizing constant.  

\item[Generalized inverse Gaussian distribution $(p, q)$] For $\lambda \in \R$, $a,b >0$, the \emph{generalized inverse Gaussian} distribution with parameters $(\lambda,a,b ; p, q)$, which we denote $\mathrm{GIG} (\lambda,a,b ; p, q)$, has density
\[\frac{1}{Z}x^{\lambda-1}e^{-a  p  x-bq x^{-1}},\qquad x\in\mathbb{R}_+,\]
where $Z$ is a normalizing constant.  
\end{description}

We also list special classes of them, which are more common in literature. In the following, $\mathrm{Be}'(\lambda,a,b ; p, q)$, $\mathrm{K}(\lambda,a,b ; p, q)$ and $\mathrm{GIG}(\lambda,a,b ; p, q)$ are extend properly for $p, q \in \{0,\infty\}$ by the natural way. 

\begin{description}
\item[Generalized Beta prime distribution $(\delta)$]  For $a,b,\delta>0$ and $c \in \R$, the \emph{Generalized Beta prime} distribution with parameters $(a,b,c ; \delta)$, which we denote $\mathrm{Be}'_{\delta}(a, b, c)$, has density
\[\frac{1}{Z} x^{a-1} (1+x)^{-a-b} \left(\frac{1+\delta x}{1+x} \right)^{c},\qquad x\in\mathbb{R}_+,\]
where $Z$ is a normalizing constant. Note that $\mathrm{Be}'(\la,a,b ; \delta, 1)=\mathrm{Be}'_{\delta}(b+\frac{\la}{2}, a-\frac{\la}{2}, -a-\frac{\la}{2})$. 
\item[Beta prime distribution]  For $a,b>0$, the \emph{Beta prime} distribution with parameters $(a,b)$, which we denote $\mathrm{Be}'(a,b)$, has density
\[\frac{1}{Z} x^{a-1} (1+x)^{-a-b} ,\qquad x\in\mathbb{R}_+,\]
where $Z$ is a normalizing constant. Note that $\mathrm{Be}'(\la,a,b ; 1,0)= \mathrm{Be}'(\la, a-\frac{\la}{2})$ for $\la >0$, $\mathrm{Be}'(\la,a,b ; 0,1)= \mathrm{Be}'(b+\frac{\la}{2}, -\la)$ for $\la<0$, $\mathrm{K}(\la,a,b ; 0,1)= \mathrm{Be}'(b+\frac{\la}{2}, -\la)$ for $\la <0$,  and $\mathrm{Be}'_{1}(a, b, c)= \mathrm{Be}'(a, b)$.
\item[Kummer distribution of Type 2]  For $a,c>0$ and $b \in \R$, the \emph{Kummer} distribution with parameters $(a,b,c)$, which we denote $\mathrm{K}^{(2)}(a, b, c)$, has density
\[\frac{1}{Z} x^{a-1} (1+x)^{-a-b} e^{-cx},\qquad x\in\mathbb{R}_+,\]
where $Z$ is a normalizing constant. Note that $\mathrm{K}(\la, a,b ; p,1)=\mathrm{K}^{(2)}(b+\frac{\la}{2}, -\la, ap)$. 
\item[Gamma distribution]  For $\la,a>0$, the \emph{Gamma} distribution with parameters $(\la,a)$, which we denote $\mathrm{Ga}(\la, a)$, has density
\[\frac{1}{Z} x^{\la-1}e^{-ax},\qquad x\in\mathbb{R}_+,\]
where $Z$ is a normalizing constant. Note that $\mathrm{K}(\la, a,b ; p, 0)=\mathrm{Ga}(\la,ap)$ for $\la>0$, $\mathrm{K}(\la,a,b ; p, \infty)=\mathrm{Ga}(b+\frac{\la}{2},ap)$ and $\mathrm{GIG}(\la, a,b ; p, 0)= \mathrm{Ga}(\la,ap)$ for $\la>0$.
\item[Generalized inverse Gaussian distribution]  For $\la \in \R$ and $a,b>0$, the \emph{generalized inverse Gaussian}  with parameters $(\la,a,b)$, which we denote $\mathrm{GIG}(\la, a,b)$, has density
\[\frac{1}{Z} x^{\la-1}e^{-ax-bx^{-1}},\qquad x\in\mathbb{R}_+,\]
where $Z$ is a normalizing constant. Note that $\mathrm{GIG}(\la,a,b ; p, q)=\mathrm{GIG}(\la,ap,bq)$.
\end{description}

We also note that the class of distributions $\mathrm{K} (\lambda,a,b ; p, q)$ is called Whittaker distribution in \cite{Omair}. 

\section{Yang-Baxter maps on $\mathbb{CP}^1$ and $\R_+$}\label{sec:YB-map}
In this section, we briefly review the background on the Yang-Baxter maps and known results on the classification of Yang-Baxter maps on $\mathbb{CP}^1 \times \mathbb{CP}^1$ in a special class of functions, called quadrirational functions. Then, we discuss when we can restrict these Yang-Baxter maps to the domain $\R_+ \times \R_+$, which is the case we are interested in. 

The Yang-Baxter maps, whose notion was introduced in \cite{D} and the term was proposed in \cite{Ves}, are bijective maps of a Cartesian product of two identical sets $\mathcal{X}$
\[
F : \mathcal{X} \times  \mathcal{X} \to \mathcal{X} \times  \mathcal{X}
\]
satisfying the \lq\lq set-theoretical" Yang-Baxter equation
\[
F_{12} \circ F_{13} \circ F_{23} = F_{23} \circ F_{13} \circ F_{12}. 
\]
Here, $F_{ij}$ are maps on the product of three sets $\mathcal{X} \times \mathcal{X} \times \mathcal{X}$ into itself and act as $F$ on the $i$-th and $j$-th factors and as the
identity on the other. 
For a set of bijective maps $F(\a,\b) : \mathcal{X} \times  \mathcal{X}\to \mathcal{X}\times  \mathcal{X}$ with parameters $\a,\b$ in a certain set of parameters $\Theta$, we also say that they are Yang-Baxter maps if
\begin{equation}\label{eq:YB}
F_{12}(\la_1,\la_2) \circ F_{13}(\la_1,\la_3) \circ F_{23}(\la_2,\la_3) = F_{23}(\la_2,\la_3) \circ F_{13}(\la_1,\la_3) \circ F_{12}(\la_1,\la_2) 
\end{equation}
holds for any parameters $\la_1,\la_2$ and $\la_3 \in \Theta$. Actually, by replacing $\mathcal{X}$ with $\mathcal{X} \times \Theta$ and considering 
\[
\tilde{F} ((x,\a), (y,\b)): =F(\a, \b)(x,y),
\]
we obtain a (parameter-independent) Yang-baxter map $\tilde{F}$. 

Many important examples of the Yang-Baxter maps were found in literature and it is not possible to enumerate them all, but some examples are in \cite{H1997,E2003} and some classification results for the case when $\mathcal{X}$ is a finite set were obtained in \cite{Lu2000, E1999}. For more background on the Yang-Baxter maps and its transfer dynamics, see the nice review paper \cite{Ve2007}. 

In \cite{Adler}, a classification of quadrirational maps on $\mathbb{CP}^1 \times \mathbb{CP}^1$ is given and it is shown that any quadrirational map is equivalent, by some M\"obius transformations acting independently on each variable, to some Yang-Baxter map. To state their result more precisely, recall that a bijection 
\[
F : \mathbb{CP}^1 \times \mathbb{CP}^1 \to \mathbb{CP}^1 \times \mathbb{CP}^1, \quad (x,y) \mapsto (u(x,y),v(x,y))
\]
is said to be birational if $F$ and $F^{-1}$ are both rational functions. The authors of \cite{Adler} defined that a map $F : \mathbb{CP}^1 \times \mathbb{CP}^1 \to \mathbb{CP}^1 \times \mathbb{CP}^1$ is said to be \textit{quadirational} if $F$ and $\bar{F}$, which is called the companion map of $F$, satisfying $\bar{F}(u,y)=(x,v)$ for $(u,v)=F(x,y)$ is a well-defined bijection on $\mathbb{CP}^1 \times \mathbb{CP}^1$ and moreover, $F$ and $\bar{F}$ are both birational functions. In other words, $F$ is quadrirational if $F, F^{-1}, \bar{F}$ and $\bar{F}^{-1}$ are well-defined rational functions. They proved that quadrirational maps $F(x,y)=(u(x,y),v(x,y))$ have the form:
\[
u(x,y)=\frac{a(y)x+b(y)}{c(y)x+d(y)}, \quad v(x,y)=\frac{A(x)y+B(x)}{C(x)y+D(x)}
\]
where $a(y),\dots,d(y)$ are polynomials in $y$ and $A(x),\dots,D(x)$ are polynomials in $x$, whose degrees are all less than or equal to two. Hence, there exist three subclasses of such maps, which are denoted by
pair of numbers as $[1:1], [1:2]$ and $[2:2]$ depending on the highest degrees of the coefficients of the polynomials for $x$ and $y$. The most rich and interesting subclass is $[2:2]$, and so the case is studied in detail in \cite{Adler} and also in \cite{PSTV}. 

To classify quadrirational maps, the authors of \cite{Adler} introduce the equivalence with respect to M\"obius transformations acting independently on each variable $x,y,u,v$ : 

\begin{df}
Quadrirational maps $F$ and $\tilde{F}$ are (M\"ob)-equivalent if there exist M\"obius transformations  $g_1,g_2,h_1,h_2$ on $\mathbb{CP}^1$ such that
\[
\tilde{F}  = (h _1 \times h _2) \circ F  \circ (g _1 \times g_2),
\]
namely 
\[
\tilde{F} (x,y) = \left(h_1(u(g_1(x),g_2(y))), h_2 (v(g_1(x), g_2(y))) \right)
\]
where $F(x,y)=(u(x,y),v(x,y))$.
\end{df}
Theorem 1 of \cite{Adler} states that, in the subclass $[2:2]$ of quadrirational maps, up to this (M\"ob)-equivalence, there are only five families of quadrirational maps $F_I=(F_I^{\a,\b}), F_{II}=(F_{II}^{\a,\b}), \dots, F_{V}=(F_{V}^{\a,\b})$ where each of them has two complex parameters $\a, \b$. Remarkably, they found that all of these five canonical representative maps are Yang-Baxter maps, and moreover involutions and coincide with their companion maps. 

However, \cite{PSTV} pointed out that not all quadrirational maps satisfy the Yang--Baxter relation, since the M\"obius transformations on each variable, in general, destroy the Yang-Baxter property. Hence, the classification result was further refined in \cite{PSTV}, considering the following equivalence. 

\begin{df}
Families of parameter dependent quadrirational maps $F=(F^{\a,\b})_{\a, \b}$ and $\tilde{F}=(\tilde{F}^{\a,\b})_{\a, \b}$ are YB-equivalent if there exists a family of bijections $\phi(\a) : \mathbb{CP}^1 \to \mathbb{CP}^1$ such that
\[
\tilde{F}^{\a,\b} = (\phi(\a)^{-1} \times \phi(\b)^{-1}) \circ F^{\a,\b} \circ (\phi(\a) \times \phi(\b)).
\]
\end{df}
Proposition 1 of \cite{PSTV} showed that if $F=(F^{\a,\b})_{\a, \b}$ and $\tilde{F}=(\tilde{F}^{\a,\b})_{\a, \b}$ are YB-equivalent, then $F$ satisfies the relation \eqref{eq:YB} if and only if $\tilde{F}$ satisfies it. Hence, this is a natural equivalence among the parameter-dependent Yang-Baxter maps. Theorem 2 of \cite{PSTV} showed that, up to this YB-equivalence, there are 10 families of quadrirational Yang-Baxter maps of subclass $[2:2]$: the families $F_I, F_{II}, \dots, F_{V}$ obtained in \cite{Adler} and additional 5 families named $H_I, H_{II}, H_{III}^A, H_{III}^B$ and $H_V$, where all of them also have two complex parameters $\a, \b$. For $* \in \{I,II,V\}$, $F_{*}$ and $H_{*}$ are (M\"ob)-equivalent but not YB-equivalent respectively, and $F_{III}, H_{III}^A, H_{III}^B$ are (M\"ob)-equivalent but not YB-equivalent.

As noted in \cite{PSTV}, $H_I$ and $H_{II}$ have convenient subtraction-free representatives, which are $H_{I}^+$ and $H_{II}^+$ given in Introduction, and $H_{III,A}, H_{III,B}$ are originally subtraction-free. Hence, we can consider them as Yang-Baxter maps on $\R_+$ by restricting the domain and the codomain to $\R_+ \times \R_+$ and parameters $\a,\b$ to be positive. Namely, these four families are natural candidate to study the IP property. 

Since $H_{III,B} =F_{\mathrm{GIG}}$, the IP property for $H_{III,B}$ was already shown in \cite{CSirf}. Moreover, the (M\"ob)-equivalence preserves the IP property which means that $H_{III}^A$ would also have the IP property. In fact, by a direct computation, we have
\[
H_{III,B} = ( (I \circ \theta_{\a}) \times I_d) \circ H_{III,A}  \circ (  I_d  \times ( I \circ \theta_{\b}) )
\]
where $I_d$ is the identity map, and by this relation, we can easily conclude the claim of Theorem \ref{thm:ip} (iii). Hence, the essential novelty of Theorem \ref{thm:ip} is in the claims for $H^+_I$ and $H^+_{II}$. 

In \cite{PSTV}, it is also mentioned that there is an obvious (singular) limit procedure starting from $H_I^+$ to obtain $H_{II}^+$ and $H_{III}^A$. This is useful to understand relations between probability measures appeared in Theorem \ref{thm:ip}, so we state this relation more explicitly as follows :
\begin{lem}\label{lem:scale}
The following scaling limits hold in the sense of the pointwise convergence.
\begin{align*}
(i) \ & \lim_{\epsilon \downarrow 0} \ H^{+,\epsilon\a, \epsilon\b}_{I} = H^{+,\a, \b}_{II}. \\ 
(ii) \ & \lim_{\epsilon \downarrow 0} \ (\theta_{\epsilon} \times \theta_{\epsilon}) \circ H^{+,\a, \b}_{II} \circ (\theta_{\epsilon} \times \theta_{\epsilon})^{-1}= H^{\a, \b}_{III,A}.
\end{align*}
\end{lem}

\section{Independence preserving property}\label{sec:IP}
In this section, we introduce a setting in which we study the independence preserving property. Then, we list specific functions that are already known to have this property, in the historical order with a brief background. We have not found any literature that discusses functions having the IP property in a unified manner, and this section is the first such attempt.

\subsection{The independence preserving property}\label{subsec:setting}

Let $\mathfrak{F}$ be the set of measurable bijections between two product of open intervals of $\R$ :
\[
\mathfrak{F} :=\{ F : I_1 \times I_2 \to J_1 \times J_2  \ | \ I_1,I_2, J_1,J_2 : \text{open intervals of} \ \R, \  F : \text{measurable bijection}  \}.
\]
For $F \in \mathfrak{F}$, we are interested in the existence and the characterization of independent random variables $X,Y$ taking values in $I_1, I_2$ respectively such that $U,V$ are also independent with $(U,V):=F(X,Y)$. If there exist such pair of non-constant random variables $X,Y$, we say that $F$ has the independence preserving property. Studying about this IP property is equivalent to study the existence and the characterization of quadruplets of probability measures $\mu,\nu,\tilde{\mu},\tilde{\nu}$ with suitable supports, which are not delta measures, satisfying $\tilde{\mu} \times \tilde{\nu}=F(\mu \times \nu)$. In \cite{CSirf}, the relation $\tilde{\mu} \times \tilde{\nu}=F(\mu \times \nu)$ is called the detailed balance equation, and the quadruplets of probability measures $\mu,\nu,\tilde{\mu},\tilde{\nu}$ are called solutions of the detailed balance equation, so we follow this terminology. Solutions of the detailed balance equation are also called as solutions for short in the following. 

To study this question, it is natural to introduce an equivalence relation among such bijections. 
\begin{df}\label{df:IPeq}
For $F : I_1 \times I_2 \to J_1 \times J_2$ and $\tilde{F} : \tilde{I_1} \times \tilde{I_2} \to \tilde{J_1} \times \tilde{J_2}$ in $\mathfrak{F}$, we denote $F \sim \tilde{F}$ if one of the followings hold:\\
(a)  There exist four measurable bijections $g_1,g_2,h_1,h_2$ such that
\[
g_i : \tilde{I_i} \to I_i, \quad  h_ i : J_i \to \tilde{J_i}, \quad i=1,2,
\]
and
\[
\tilde{F} = (h_1 \times h_2) \circ F \circ (g_1 \times g_2).
\]
(b) $\tilde{F}=F^{-1}$.\\
(c) $\tilde{F}=F \circ \pi$ where $\pi(x,y)=(y,x)$.\\
Moreover, $F$ and $\tilde{F} \in \mathfrak{F}$ are IP-equivalent if there exists a finite sequence $F_0=F, F_1,F_2,\dots,F_n =\tilde{F} \in \mathfrak{F}$ such that $F_i \sim F_{i+1}$ for $i=0,1,\dots,n-1$. 
\end{df}
The condition (a) means that $\tilde{F}$ is obtained from $F$ by a certain coordinate-wise change of variables, which is similar to the (M\"ob)-equivalence. It is obvious that if $F$ and $\tilde{F}$ are IP-equivalent, then $F$ has the IP property if and only if $\tilde{F}$ has the IP property. In particular, without loss of generality, we can only consider functions $F:\R_+^2 \to \R_+^2$ or $F: \R^2 \to \R^2$, or fix any intervals as domains and codomains. However, it should be noted that by the change of variables, the probability distributions satisfying the IP property, namely the solutions of the detailed balance equation, also change. In particular, up to this IP equivalence, in general (more precisely the case where $\mu$ and $\nu$ are continuous distributions with supports $I_1$ and $I_2$ respectively), we can change the solution to be the product of uniform probability measures on $[0,1]$ by using the distribution functions of $\mu$ and $\nu$ for the coordinate-wise change of variables. Hence, when we apply the IP property to characterize some probability distributions, the choice of representative $F$ plays an essential role. As we see in the next subsection, all of known examples have birational representatives or piecewise linear representatives and such representatives characterize important probability distributions, such as normal, gamma, exponential, beta and so on.

\begin{rem}
The IP property is also know to hold for $F: \mathcal{X} \times \mathcal{X}  \to \mathcal{X} \times \mathcal{X}$ where $\mathcal{X}$ are discrete sets (cf. \cite{CS,CSirf}) or a set of positive definite matrices with size $r$ (cf. \cite{LW, LW2}). In this paper, we focus on the case where we can reduce the problem to $\mathcal{X}=\R_+$.
\end{rem}

\subsection{History on the study of bijections with the IP property}

The IP property was first discovered for the bijection 
\[
F_{\mathrm{N}}(x,y)=\left( x+y, \ x-y \right) : \R^2 \to \R^2.
\] 
The solution of the detailed balance equation for this bijection was characterized independently by Kac \cite{Kac} (1939) and Bernstein \cite{Be} (1941) as $X \sim \mathrm{N}(a, \sigma), Y \sim \mathrm{N}(b, \sigma)$ for any $a,b \in \R$ and $\sigma >0$ where $\mathrm{N}(a,\sigma)$ is the normal distribution with the mean $a$ and the variance $\sigma$. The original characterization results were given under some integrability assumption on the distributions, but later the characterization was given in full generality (cf. \cite{KLR}). 

After this impressive discovery, several bijections having the IP property have been introduced in different contexts. Most classical ones were found in the context of characterization of important probability distributions. Such examples are 
\[
F_{\mathrm{Ga}}(x,y)=\left(x+y, \ \frac{x}{y} \right) : \R_+^2 \to \R_+^2,  \quad F_{\mathrm{Exp}}(x,y)=(\min\{x,y\}, \ x-y) : \R^2 \to \R^2.
\] 
The solutions for $F_{\mathrm{Ga}}$ were characterized by Lucaks \cite{L} (1955) as $X \sim \mathrm{Ga}(a,\la)$, $Y \sim \mathrm{Ga} (b,\la)$ for any $a,b,\la >0$. For $F_{\mathrm{Exp}}$, the solutions were characterized by Ferguson \cite{Fe1,Fe2} (1964,1965) as pairs of (possibly shifted) Exponential distributions or (possibly shifted) Geometric distributions with certain parameters. Hence, the IP property for each bijection characterizes Gamma, Exponential and Geometric distributions. Note that $F_{\mathrm{Exp}}$ can be considered as a zero-temperature limit of $F_{\mathrm{Ga}}$, namely $F_{\mathrm{Exp}}$ is obtained from $F_{\mathrm{Ga}}$ by replacing $(+,\times)$-algebra with $(\min,+)$-algebra. Consistent with this, Exponential distributions can be understood as a zero-temperature limit of Gamma distributions as discussed in \cite{CSirf}. In particular, as discussed in \cite{CSirf}, the IP property is typically inherited to the zero-temperature limit, and the zero-temperature limit version has both continuous and discrete solutions. 

As a slightly different example, the function 
\[
F_{\mathrm{C}}(x,y)=\left(y, \ \frac{x+y}{1-xy} \right) : \R^2 \setminus \{(x,y) \in \R^2  \ | \ xy=1 \} \to \R^2
\]
is known to have the IP property which characterizes the Cauchy distribution. Precisely, though $F_{\mathrm{C}}$ is not well-defined on $\R^2$, when $X,Y$ are independent absolutely continuous random variables, then $(U,V)$ is well-defined almost surely and under this condition, $(U,V)$ are independent if and only if $X$ is the Cauchy distribution with a specific parameter, which was shown in \cite{Arnold} (1979). This case, $Y$ can be any absolutely continuous random variable. Moreover, by change of variables $x \to \arctan x$ and $y \to \arctan y$, the IP property for the equivalent function is also known to characterize the uniform distribution. 

After a blank period around twenty years, another bijection 
\[
F_{\mathrm{Be}}(x,y) =\left(\frac{1-y}{1-xy}, 1-xy \right) : (0,1)^2 \to (0,1)^2
\]
was found to have the IP property in \cite{W2003} (2003) and the solutions were completely characterized in \cite{SW} as $X \sim \mathrm{Be}(a,b)$, $Y \sim \mathrm{Be}(a+b,c)$ for any $a,b,c >0$ where $\mathrm{Be}(a,b)$ is the Beta distribution with shape parameters $a,b$. 

From a very different context, which was a study of an exponential-version of Pitman's transform for geometric Brownian motions, Matsumoto and Yor considered the involution 
\[
F_{\mathrm{GIG-Ga}}(x,y)=\left(\frac{1}{x+y}, \ \frac{1}{x}-\frac{1}{x+y}\right) =\left(\frac{1}{x+y}, \ \frac{y}{x(x+y)}\right) : \R_+^2 \to \R_+^2
\] 
in \cite{MY2} (see also \cite{MY, MY1}) (2001) and found that this function has the IP property with $X \sim \mathrm{GIG}(-\la,a,b)$ and $Y \sim \mathrm{Ga}(\la,a)$ for $\la,a, b>0$. The complete characterization of the solutions was done in \cite{LW} (2000). Then, to generalize this result, Koudou and Vallois \cite{KV} (2012) considered the class of bijections of the form 
\begin{equation}\label{MY}
F(x,y)=\left(f(x+y), \ f(x)-f(x+y)\right) : \R_+^2 \to \R_+^2
\end{equation}
given by a decreasing three times differentiable bijection $f : \R_+ \to \R_+$. Remarkably, they completely characterized bijections $F$ in this class having the IP property for some probability distributions with positive and twice differentiable densities. Up to the IP-equivalence we introduced, they proved that other than the  $f(x)=\frac{1}{x}$ which leads $F_{\mathrm{GIG-Ga}}$, there are only two cases having the IP property : $f(x)=\log\left(\frac{1+x}{x}\right)$ or $f(x)=\log \left( \frac{e^x+\delta-1}{e^x-1} \right)$ with a parameter $\delta >0$. By introducing a certain coordinate-wise change of variables, they showed that $f(x)=\log\left(\frac{1+x}{x}\right)$ leads the bijection
\[
F_{\mathrm{K-Ga},A}(x,y)= \left( x+y, \ \frac{1+\frac{1}{x+y}}{1+\frac{1}{x}}\right) : \R_+^2 \to \R_+ \times (0,1)
\]
and $f(x)=\log \left( \frac{e^x+\delta-1}{e^x-1} \right)$ leads the involution
\[
F_{\mathrm{Be}}^{\delta}(x,y) = \left(\frac{1-xy}{1+ (\delta-1) xy}, \ \frac{1-x}{1+ (\delta-1) x} \frac{1+ (\delta-1) xy}{1-xy} \right) : (0,1)^2 \to (0,1)^2
\]
with a parameter $\delta >0$. When $\delta=1$, $F_{\mathrm{Be}}^{1}$ is IP-equivalent to $F_{\mathrm{Be}}$ since $F_{\mathrm{Be}}= \pi \circ F_{\mathrm{Be}}^{1} \circ \pi$. The solutions for $F_{\mathrm{K-Ga},A}$ are given as $X \sim \mathrm{K}^{(2)}(a,b,c)$ and $Y \sim \mathrm{Ga}(b,c)$ for $a, b,c >0$ while those for $F_{\mathrm{Be}}^{\delta}$ are given as $X \sim \mathrm{Be}_{\delta}(a+b,\la, -\la-b)$ and $Y \sim \mathrm{Be}(a,b)$ for $a, b,\la >0$ where $\mathrm{Be}_{\delta}$ is related by the change of variable $x \to \frac{x}{1-x} : (0,1) \to \R_+$ to $\mathrm{Be}'_{\delta}$. Note that $F_{\mathrm{Ga}}$ is IP-equivalent to a bijection of the form \eqref{MY} with $f(x)=\log x$, but since this $f$ takes negative values, it was not included in the framework of \cite{KV}. The functions of the form \eqref{MY} having the IP property are also said to have Matsumoto-Yor property. 

In \cite{HAMZA} (2015), the authors introduced another bijection with the IP property
\[
F_{\mathrm{K-Ga},B}(x,y) = \left(\frac{y}{1+x}, \ \frac{x(1+x+y)}{1+x} \right) : \R^2_+ \to \R^2_+
\] 
which also involves Kummer distribution and Gamma distribution with a motivation to give a closed identity satisfied by the Kummer distribution. 
The characterization of solutions without any assumption on distributions for $F_{\mathrm{K-Ga},A}$ and $F_{\mathrm{K-Ga},B}$ are given in \cite{PW} (2018). 

Most recently, in the study of invariant measures for the discrete KdV equation, which is a well-known discrete integrable system, one of the authors of the present paper found, with the coauthor of \cite{CSirf}, that for the class of involutions $F_{\mathrm{GIG}}^{\a,\b}$, the IP property holds with $X \sim \mathrm{GIG}(\la, a \a, b)$ and $Y \sim \mathrm{GIG}(\la, b \b,a)$ for $\la \in \R, a,b>0$. Moreover, the special case $\a=1,\b=0$ is IP-equivalent to $F_{\mathrm{GIG-Ga}}$ since 
\[
F_{\mathrm{GIG-Ga}} = (I_d \times I) \circ F_{\mathrm{GIG}}^{1,0} \circ  (I_d \times I). 
\]
The zero-temperature version of $F_{\mathrm{GIG}}^{\a,\b}$, which is related to the ultra-discrete KdV equation, and the zero-temperature version of $F_{\mathrm{Be}}$ are also introduced in the study of discrete integrable systems and stochastic integrable systems, and shown to have the IP property \cite{CSirf, CSjsp}:
\begin{align*}
& F^{\a,\b}_{\mathrm{GIG,zero}}(x,y) \\
& = (x+\min\{y, \a-x\} - \min\{x, \b-y\}, \ y-\min\{y, \a-x\} + \min\{x, \b-y\} ) : \R^2 \to \R^2\\
&  F_{\mathrm{Be,zero}}(x,y)   =(\min\{x, 0\} -y, \ \min\{x,y,0\}-x-y)  : \R^2 \to \R^2.
\end{align*}
The characterization of solutions for them are studied in \cite{CSirf, BN}, but not fully solved yet, even for the special case $(\a,\b)=(0,\infty)$:
\[
F^{0,\infty}_{\mathrm{GIG,zero}}(x,y) = (\min\{y, -x\} , \ y+x -\min\{y, -x\}  )  : \R^2 \to \R^2
\]
which is IP-equivalent to the zero-temperature version of $F_{\mathrm{GIG-Ga}}$. 

As far as we are aware, there is no other known (nontrivial) functions on the two product of open intervals of $\R$ having the IP property. In the next subsection, we discuss common properties of them and give a classification. 

\subsection{Normalized bijections having the IP property in literature and its classification}

The bijections introduced in the last subsection are simply classified into two classes : the zero-temperature version of some other bijection, and the rest of them. All of the zero-temperature versions, namely $F_{\mathrm{Exp}}$, $F^{\a,\b}_{\mathrm{GIG,zero}}$ and $F_{\mathrm{Be,zero}}$ are piecewise linear functions and involve the \lq\lq $\min$" function. Since the IP property of these zero-temperature versions can be understood from the same property of the original (namely, the corresponding \lq\lq positive-temperature") bijection, from now on, we focus on the rest of them, namely, $F_{\mathrm{N}}$, $F_{\mathrm{Ga}}$, $F_{\mathrm{C}}$, $F_{\mathrm{Be}}$, $F_{\mathrm{GIG-Ga}}$, $F_{\mathrm{K-Ga},A}$, $F_{\mathrm{Be}}^{\delta}$, $F_{\mathrm{K-Ga},B}$, $F_{\mathrm{GIG}}^{\a,\b}$ and consider the relation between them. 

\begin{rem}
As mentioned in \cite{PSTV} and already discussed in Introduction, there should be natural zero-temperature versions of $H_I^+$ and $H_{II}^+$. Applying Theorem \ref{thm:ip}, we will be able to show that the zero-temperature versions of $H_I^+$ and $H_{II}^+$ have the IP property and their solutions should be the zero-temperature version of generalized Beta prime distributions and Kummer distributions of Type $2$. 
\end{rem}

From the explicit expressions of these bijections, we can find that they are all birational and naturally extended to the birational functions on $\mathbb{CP}^1 \times \mathbb{CP}^1$. Hence, we may consider coordinate-wise M\"obius transformations to normalize these functions to compare. Namely, instead of considering the equivalence with respect to (a) of Definition \ref{df:IPeq}, consider the (M\"ob)-equivalence. Then, any open interval $I \subsetneq \R$ can be mapped to $\R_+$ by a M\"obius transformation, but $\R$ cannot. Hence, in this sense, $F_{\mathrm{N}}$ and $F_{\mathrm{C}}$ are exceptional, which are not (M\"ob)-equivalent to a birational function which can be restricted to the one having domains and codomains $\R_+^2$. In other words, $F_{\mathrm{Ga}}, F_{\mathrm{Be}},F_{\mathrm{GIG-Ga}}, F_{\mathrm{K-Ga},A}, F_{\mathrm{Be}}^{\delta}$, $F_{\mathrm{K-Ga},B}$, $F_{\mathrm{GIG}}^{\a,\b}$ share the common property that they are (M\"ob)-equivalent to a subtraction-free birational function with domains and codomains $\R_+^2$. Hence, to study relations between these bijections, we give such representatives for them, which are denoted by $F_*^+$, explicitly by applying the change of variable $x \to \frac{x}{1-x} : (0,1) \to \R_+$ for each coordinate in $F_{\mathrm{Be}}^{\delta}$ and the second component of the output of $F_{\mathrm{K-Ga},A}$ :
\begin{itemize}
\item $F^+_{\mathrm{Ga}}(x,y)=F_{\mathrm{Ga}}(x,y) = \left(x+y, \frac{x}{y} \right), \ (F^+_{\mathrm{Ga}})^{-1}(x,y)= \left( \frac{xy}{1+y}, \frac{x}{1+y} \right)$
\item $F^{+,\delta}_{\mathrm{Be}}(x,y)= \left(\frac{1+x+y}{\delta xy },  \ \frac{1+x+y+\delta xy}{x(\delta +\delta x)} \right), \ (F^{+,\delta}_{\mathrm{Be}})^{-1}=F^{+,\delta}_{\mathrm{Be}}$
\item $F^+_{\mathrm{K-Ga},A}(x,y)=\left(x+y, \frac{x(x + y +1)}{y} \right), \ (F^+_{\mathrm{K-Ga},A})^{-1}(x,y)=\left(\frac{xy}{1+x+y}, \frac{x(1+ x)}{1+ x+y} \right)$
\item $F^+_{\mathrm{K-Ga},B}(x,y)=F_{\mathrm{K-Ga},B}(x,y) =\left(\frac{y}{1+x}, \frac{x(1+x+y)}{1+x} \right), \ (F^+_{\mathrm{K-Ga},B})^{-1}=F^+_{\mathrm{K-Ga},B}$
\item $F^{+,\a,\b}_{\mathrm{GIG}}(x,y)=F^{\a,\b}_{\mathrm{GIG}}(x,y)=\left(y \frac{1+\b xy }{1+ \a xy},  \ x \frac{1+\a xy }{1+ \b xy} \right), \ (F^{+,\a,\b}_{\mathrm{GIG}})^{-1}=F^{+,\a,\b}_{\mathrm{GIG}}$
\end{itemize}
For $F_{\mathrm{Be}}$ and $F_{\mathrm{GIG-Ga}}$, since we know they are (M\"ob)-equivalent to $F^{1}_{\mathrm{Be}}$ and $F^{1,0}_{\mathrm{GIG}}$ respectively, we choose representatives as 
\begin{itemize}
\item $F^{+}_{\mathrm{Be}}(x,y)=F^{+,1}_{\mathrm{Be}}(x,y) =\left(\frac{1+x+y}{xy },  \ \frac{1+y}{x} \right)$
\item $F^+_{\mathrm{GIG-Ga}}=F^{+,1,0}_{\mathrm{GIG}}(x,y)=\left(\frac{y }{1+ xy},  \ x (1+ xy ) \right)$
\end{itemize}
Other than these two simple equivalence, any relation between these bijections are not discussed in literature. From these explicit expressions, we can easily see that some of them are obtained by a singular limit procedure from others by introducing scaling parameters. Such procedure conserves the IP property, and so the IP property of them are also related. Surprisingly, all of these bijections are obtained from $H_I^{+}, H_{II}^+$ and $H_{III,A}$ by such procedure and/or taking special values of parameters together with a proper coordinate-wise change of variables, which is our second main result stated in Theorem \ref{thm:relation} in Section \ref{sec:special}. 

\section{Proof of Theorem \ref{thm:ip} and properties of probability distributions}\label{sec:proof}

In this section, we discuss proofs of Theorem \ref{thm:ip}. In the first subsection, we give a proof by a direct computation for the most complicated case, namely for the claim (i). In the second subsection, we study some useful properties of probability distributions appearing in Theorem \ref{thm:ip}, and explain a way to obtain Theorem \ref{thm:ip} (ii) and (iii) from Theorem \ref{thm:ip} (i). 

\subsection{Proof of Theorem \ref{thm:ip}}
In this subsection, we give a proof of Theorem \ref{thm:ip}. Actually, the result is given by a direct computation and can be checked even by computers, but for clarity, we give some key formulas. 

To prove the claims of Theorem \ref{thm:ip}, we only need to prove that 
\[
p_X(x)p_Y(y)=J_F(x,y)p_U(u(x,y))p_V(v(x,y))
\]
holds where $F(x,y)=(u(x,y),v(x,y))$ and $J_F(x,y)$ is the Jacobian of $F$, namely $J_F(x,y)=|\frac{\partial u}{\partial x} \frac{\partial v}{\partial y} - \frac{\partial v}{\partial x} \frac{\partial u}{\partial y}|$ and $p_X,p_Y,p_U,p_V$ are probability density functions for the distributions and parameters given each claim. 

To check the relation for (i), let $u(x,y)=\frac{y}{\a}\frac{\sigma_1(x,y)}{\sigma_2(x,y)}$ and $v(x,y)=\frac{x}{\b} \frac{\sigma_3(x,y)}{\sigma_4(x,y)}$, namely 
\begin{align*}
&\sigma_1(x,y)=\b+\a x+\b y + \a \b xy, &\sigma_2(x,y)=1+x+y+\b xy,  \\
& \sigma_3(x,y)=\a+\a x+\b y + \a \b xy, &\sigma_4(x,y)=1+x+y+\a xy.
\end{align*}
Then, by simple calculations, we have
\begin{align*}
&  y \sigma_1 +  \sigma_2 = \sigma_4 (1+\b y),  \quad   y \sigma_1 + \a \sigma_2  = \sigma_3 (1+y), \\
& x \sigma_3 + \sigma_4 = \sigma_2 (1+\a x),  \quad   x \sigma_3 + \b \sigma_4 = \sigma_1 (1+x), \\
& J_F(x,y) = \frac{\sigma_1 \sigma_3}{\a \b \sigma_2 \sigma_4}
\end{align*}
where we simply denote $\sigma_i(x,y)$ by $\sigma_i$ for $i=1,\dots,4$. Hence, if $U \sim \mathrm{Be}' (-\la, a, b ; \a ,1)$, we have 
\begin{align*}
p_U(u(x,y))& = \frac{1}{Z} \left(\frac{y}{\a}\frac{\sigma_1}{\sigma_2} \right)^{-\lambda-1} \left(1+y\frac{\sigma_1}{\sigma_2}  \right)^{-a+\frac{\lambda}{2}} \left(1+ \frac{\a}{y}\frac{\sigma_2}{\sigma_1}  \right)^{-b-\frac{\lambda}{2}} \\
& = \frac{1}{Z} y^{-\lambda-1} \frac{\sigma_1^{-\lambda-1}}{\sigma_2 ^{-\lambda-1}} \left(\frac{\sigma_2 + y\sigma_1}{\sigma_2}  \right)^{-a+\frac{\lambda}{2}} \left(\frac{y \sigma_1 + \a \sigma_2}{y \sigma_1}  \right)^{-b-\frac{\lambda}{2}} \\
& = \frac{1}{Z} y^{-\lambda-1} \sigma_1^{-\frac{\lambda}{2}-1+b} \sigma_2^{\frac{\lambda}{2}+1+a} ( \sigma_4 (1+\b y))^{-a+\frac{\lambda}{2}}  ( \sigma_3 (1+y))^{-b-\frac{\lambda}{2}}
\end{align*}
where $Z$ is a normalizing constant which may change line by line. 
In the same way, f $V \sim \mathrm{Be}' (\la, a, b ; \b ,1)$, we have 
\begin{align*}
p_V(v(x,y))& = \frac{1}{Z} \left(\frac{x}{\b}\frac{\sigma_3}{\sigma_4} \right)^{\lambda-1} \left(1+x\frac{\sigma_3}{\sigma_4}  \right)^{-a-\frac{\lambda}{2}} \left(1+ \frac{\b}{x}\frac{\sigma_4}{\sigma_3}  \right)^{-b+\frac{\lambda}{2}} \\
& = \frac{1}{Z} x^{\lambda-1} \frac{\sigma_3^{\lambda-1}}{\sigma_4^{\lambda-1}} \left(\frac{\sigma_4 + x\sigma_3}{\sigma_4}  \right)^{-a-\frac{\lambda}{2}} \left(\frac{x \sigma_3 + \b \sigma_4}{x \sigma_3}  \right)^{-b+\frac{\lambda}{2}} \\
& = \frac{1}{Z} x^{\lambda-1} \sigma_3^{\frac{\lambda}{2}-1+b} \sigma_4^{-\frac{\lambda}{2}+1+a} ( \sigma_2 (1+\a x))^{-a-\frac{\lambda}{2}}  ( \sigma_1 (1+x))^{-b+\frac{\lambda}{2}}.
\end{align*}
Hence, 
\[
J_F(x,y)p_U(u(x,y))p_V(v(x,y))=\frac{1}{Z}y^{-\lambda-1}(1+\b y)^{-a+\frac{\lambda}{2}} (1+y)^{-b-\frac{\lambda}{2}} x^{\lambda-1}  (1+\a x)^{-a-\frac{\lambda}{2}} (1+x)^{-b+\frac{\lambda}{2}},
\]
which is equal to $p_X(x)p_Y(y)$ if $X \sim \mathrm{Be}' (\la, a, b ; \a ,1)$ and $Y \sim \mathrm{Be}' (-\la, a, b ; \b ,1)$. Therefore we conclude that the claim (i) holds. 

For the claim (ii) and (iii), we can prove in a similar way. In the next subsection, we give another proof by applying a singular limit procedure.

\subsection{Some properties of probability distributions under M\"obius transformations and scaling limits}

We have introduced probability distributions $\mathrm{Be}'(\lambda,a,b ; p, q)$, $\mathrm{K} (\lambda,a,b ; p, q)$ and $\mathrm{GIG} (\lambda,a,b  ; p, q)$ with two auxiliary parameters $p,q >0$ because these classes are closed under the change of scaling, and also two of them are closed under the map $I$ as follows.
\begin{lem}\label{lem:trans}
For the maps $I$ and $\theta_{\alpha}$ for any $\a >0$, the following relations hold. 

(i) Suppose $X \sim \mathrm{Be}' (\lambda,a,b ; p,q)$. Then, 
\[
I(X)=X^{-1} \sim \mathrm{Be}' (-\la, b, a ; q, p),  \quad  \theta_{\a}(X)=\alpha X  \sim \mathrm{Be}' (\la, a, b ; \frac{p}{\alpha}, \alpha q).
\]
(ii) Suppose $X \sim \mathrm{K} (\lambda,a,b ; p,q)$. Then, 
\[
\theta_{\a}(X)=\alpha X  \sim K (\la, a, b ; \frac{p}{\alpha}, \alpha q).
\]
(iii) Suppose $X \sim \mathrm{GIG} (\lambda,a,b ; p, q)$. Then, 
\[
I(X)=X^{-1} \sim GIG (-\la, b, a ; q, p),  \quad  \theta_{\a}(X)=\alpha X  \sim GIG (\la, a, b ; \frac{p}{\alpha}, \alpha q).
\]
\end{lem}
This lemma is proved by direct computations. 

Moreover, these probability measures are connected by the following scaling limits. 
\begin{lem}\label{lem:scale2}
The following scaling limits hold in the sense of weak convergence. 

(i) For $a >0$ and $b, \la \in \R$ such that $-b <\frac{\la}{2}$, 
\begin{align*}
\lim_{\epsilon \downarrow 0} \ \mathrm{Be}'(\la, \frac{a}{\epsilon}, b ; \epsilon p , q) = \mathrm{K}(\la, a, b ;  p , q). 
\end{align*}
(ii) For $a, b >0$ and $\la \in \R$,
\begin{align*}
\lim_{\epsilon \downarrow 0} \  \mathrm{K}(\la, a, \frac{b}{\epsilon} ;  p , \epsilon q)= \mathrm{GIG}(\la, a, b ;  p , q). 
\end{align*}
\end{lem}
This lemma is also proved by the explicit forms of probability density functions, by noting 
\[
\lim_{\epsilon \downarrow 0} (1+\epsilon px)^{-\frac{a}{\epsilon}-\frac{\la}{2}} =  e^{-apx}, \quad \lim_{\epsilon \downarrow 0} (1+\epsilon qx^{-1})^{-\frac{b}{\epsilon}+\frac{\la}{2}}= e^{-bqx^{-1}}.
\]
From the claim (i) of Theorem \ref{thm:ip}, for any $a >0$ and $b, \la \in \R$ such that $-b <\frac{\la}{2}$, we have the IP property holds for $H_{I}^{+, \epsilon \a, \epsilon \b}$ with 
\begin{align*}
X \sim \mathrm{Be}'  (\la, \frac{a}{\epsilon}, b \ ; \ \epsilon \a, 1), & \quad   Y \sim \mathrm{Be}'  (-\la, \frac{a}{\epsilon}, b \ ; \ \epsilon \b, 1), \\
U \sim \mathrm{Be}' (-\la, \frac{a}{\epsilon}, b \ ; \ \epsilon \a, 1), & \quad   V \sim \mathrm{Be}'  (\la, \frac{a}{\epsilon}, b \ ; \ \epsilon \b, 1)
\end{align*} 
with sufficiently small $\epsilon >0$. Hence, applying Lemmas \ref{lem:scale} (i) and \ref{lem:scale2} (i) , we conclude the claim (ii) of Theorem \ref{thm:ip}. 

By this claim (ii) of Theorem \ref{thm:ip} and Lemma \ref{lem:trans} (ii), for $(\theta_{\epsilon} \times \theta_{\epsilon}) \circ H^{+,\a,\b}_{II} \circ (\theta_{\epsilon} \times \theta_{\epsilon})^{-1}$, the IP property holds with 
\begin{align*}
X \sim \mathrm{K}  (\la, a, b \ ; \   \frac{\a}{\epsilon}, \epsilon), & \quad   Y \sim \mathrm{K}  (-\la, a, b \ ; \ \frac{\a}{\epsilon}, \epsilon), \\
U \sim \mathrm{K} (-\la, a, b \ ; \ \frac{\a}{\epsilon}, \epsilon), & \quad   V \sim \mathrm{K}  (\la, a, b \ ; \ \frac{\a}{\epsilon}, \epsilon)
\end{align*} 
By changing $a \to a \epsilon $ and $b \to \frac{b}{\epsilon}$, the IP property holds for the same function with 
\begin{align*}
X \sim \mathrm{K}  (\la, a \epsilon, \frac{b}{\epsilon} \ ; \   \frac{\a}{\epsilon}, \epsilon)= \mathrm{K}  (\la, a , \frac{b}{\epsilon} \ ; \   \a, \epsilon), & \quad   Y \sim \mathrm{K}  (-\la, a \epsilon, \frac{b}{\epsilon} \ ; \ \frac{\a}{\epsilon}, \epsilon)= \mathrm{K}  (-\la, a , \frac{b}{\epsilon} \ ; \   \a, \epsilon), \\
U \sim \mathrm{K} (-\la, a \epsilon, \frac{b}{\epsilon} \ ; \ \frac{\a}{\epsilon}, \epsilon)=\mathrm{K}  (-\la, a , \frac{b}{\epsilon} \ ; \   \a, \epsilon), & \quad   V \sim \mathrm{K}  (\la, a \epsilon, \frac{b}{\epsilon} \ ; \ \frac{\a}{\epsilon}, \epsilon)= \mathrm{K}  (\la, a , \frac{b}{\epsilon} \ ; \   \b, \epsilon)
\end{align*} 
for sufficiently small $\epsilon >0$. Then, applying Lemmas \ref{lem:scale} (ii) and \ref{lem:scale2} (ii) , we conclude the claim (iii) of Theorem \ref{thm:ip}.


\section{Reduction of known IP properties from Theorem \ref{thm:ip}}\label{sec:special}
In this section, we recover the IP property for bijections discussed in Section \ref{sec:IP} from Theorem \ref{thm:ip}. For this, we give our second main result which connect all of the known bijections having a subtraction-free representation to the newly introduced bijections $H_I^{+}, H_{II}^+$ and $H_{III,A}$. 
\begin{thm}\label{thm:relation}
The bijections $F^{+,\delta}_{\mathrm{Be}}$, $F^+_{\mathrm{K-Ga},A}$, $F^+_{\mathrm{K-Ga},B}$, $F^+_{\mathrm{Ga}}$ and $F^{+,\a,\b}_{\mathrm{GIG}}$ are obtained from one of $H_I^{+,\a,\b}, H^{+,\a,\b}_{II}$ and $H^{\a,\b}_{III,A}$ by M\"obius transformations and singular limits as follows. \\
(i) $F^{+,\delta}_{\mathrm{Be}} = \tilde{H}^{\delta,0}_{I}$ where 
\[
\tilde{H}^{\a,\b}_{I} = ( (I \circ \theta_{\a}) \times ( I \circ \theta_{\b} ) )\circ H^{+,\a,\b}_{I}.
\]
(ii) $F^+_{\mathrm{K-Ga},A} = \tilde{H}^{1,0}_{II}$ where 
\[
\tilde{H}^{\a,\b}_{II} =  H^{+,\a,\b}_{II} \circ \left(\theta_{\a^{-1}} \times \theta_{\b^{-1}}  \right).
\]
(iii) $F^+_{\mathrm{K-Ga},B} = \hat{H}^{1,0}_{II}$ where 
\[
\hat{H}^{\a,\b}_{II} =\left(\theta_{\a^{-1}} \times \theta_{\b^{-1}}  \right)  \circ  H^{+,\a^{-1},\b^{-1}}_{II} \circ \left(\theta_{\a} \times \theta_{\b}  \right).
\]
(iv) $F^+_{\mathrm{Ga}}=\tilde{F}^{0}_{\mathrm{K-Ga},A}$ where
\[
\tilde{F}^{\epsilon}_{\mathrm{K-Ga},A} = ( \theta_{\epsilon^{-1}} \times I_d) \circ  F^+_{\mathrm{K-Ga},A} \circ (\theta_{\epsilon} \times \theta_{\epsilon}) .
\]
(v) $F^+_{\mathrm{Ga}}=\tilde{F}^{0}_{\mathrm{K-Ga},B}$ where
\[
\tilde{F}^{\epsilon}_{\mathrm{K-Ga},B} = \pi \circ ( I \times \theta_{\epsilon}) \circ  F^+_{\mathrm{K-Ga},B} \circ (\theta_{\epsilon^{-1}} \times \theta_{\epsilon^{-1}}) .
\]
(vi) $F^{+,\a,\b}_{\mathrm{GIG}}=( (I \circ \theta_{\a}) \times I_d) \circ H^{\a,\b}_{III,A}  \circ (  I_d  \times ( I \circ \theta_{\b}) )$. 
\end{thm}
The theorem is proved by a direct computation. 

Applying these explicit relations and Lemma \ref{lem:trans} to Theorem \ref{thm:ip}, we have the solutions of the detailed balance equations for all of bijections appeared in the last theorem by a systematic way. Note that when parameters $p,q$ take singular values $0$ or $\infty$, the range of parameters may narrow. 
\begin{cor}\label{cor:ip}
For the following distributions $(X,Y)$, the random variables $U,V$ given by $(U,V)=F(X,Y)$ for each map are independent and have the following distributions. \\
(i) For $F=\tilde{H}^{\a,\b}_{I}$, 
\begin{align*}
X \sim \mathrm{Be}' (\la, a, b ; \a, 1), & \quad  Y \sim \mathrm{Be}' (-\la, a, b ; \b, 1), \\
U \sim \mathrm{Be}' (\la,  b,a ; \a, 1), & \quad V \sim \mathrm{Be}' (-\la, b,a  ; \b, 1)
\end{align*} 
and for $F=F^{+,\delta}_{\mathrm{Be}}$,
\begin{align*}
X \sim \mathrm{Be}'  (\la, a, b ; \delta, 1)=\mathrm{Be}'_{\delta}\left(b+\frac{\la}{2}, a-\frac{\la}{2}, -a-\frac{\la}{2} \right) ,& \quad  Y \sim \mathrm{Be}'  \left(-\la, a, b ; 0, 1 \right)=\mathrm{Be}' \left(b-\frac{\la}{2},\la \right)\\
U \sim \mathrm{Be}'  (\la,  b,a ; \delta, 1)=\mathrm{Be}'_{\delta} \left(a+\frac{\la}{2}, b-\frac{\la}{2}, -b-\frac{\la}{2} \right), & \quad V \sim \mathrm{Be}'  \left(-\la, b,a  ; 0, 1\right)=\mathrm{Be}' \left(a-\frac{\la}{2},\la \right)
\end{align*} 
where $\lambda \in \R$, $a,b >0$, $- \min\{a,b\} < \frac{\lambda}{2} < \min\{a,b\}$. \\
(ii) For $F=\tilde{H}^{\a,\b}_{II}$, 
\begin{align*}
X \sim \mathrm{K} (\la, a, b ; 1,\a), & \quad  Y \sim \mathrm{K} (-\la, a, b ; 1,\b), \\
U \sim \mathrm{K} (-\la,  a,b ; \a, 1), & \quad V \sim \mathrm{K} (\la, a,b  ; \b, 1)
\end{align*} 
where $\lambda \in \R$, $a,b >0$, $-b < \frac{\lambda}{2} < b$ and for $F=F^+_{\mathrm{K-Ga},A}$,
\begin{align*}
X \sim \mathrm{K}  (\la, a, b ; 1, 1)=\mathrm{K}^{(2)} \left(b+\frac{\la}{2},-\la,a \right) & \quad  Y \sim \mathrm{K}  (-\la, a, b ; 1,0)=\mathrm{Ga}(-\la, a)\\
U \sim \mathrm{K}   (-\la, a,b ; 1, 1)=\mathrm{K}^{(2)} \left(b-\frac{\la}{2},\la,a \right), & \quad V \sim \mathrm{K}   (\la, a,b  ; 0, 1)=\mathrm{Be}' \left(b+\frac{\la}{2},-\la \right)
\end{align*}
where $\lambda <0$, $a,b >0$, $-b < \frac{\lambda}{2}$. \\
(iii) For $F=\hat{H}^{\a,\b}_{II}$, 
\begin{align*}
X \sim \mathrm{K} (\la, a, b ; 1,\a^{-1}), & \quad  Y \sim \mathrm{K} (-\la, a, b ; 1,\b^{-1}), \\
U \sim \mathrm{K} (-\la,  a,b ; 1, \a^{-1}), & \quad V \sim \mathrm{K} (\la, a,b  ; 1,\b^{-1}).
\end{align*} 
and for $F=F^+_{\mathrm{K-Ga},B}$,
\begin{align*}
X \sim \mathrm{K}  (\la, a, b ; 1, 1)=\mathrm{K}^{(2)} \left(b+\frac{\la}{2},-\la,a \right), & \quad  Y \sim \mathrm{K}  (-\la, a, b ; 1, \infty )=\mathrm{Ga} \left(b-\frac{\la}{2}, a \right) \\
U \sim \mathrm{K}   (-\la, a,b ; 1, 1)=\mathrm{K}^{(2)} \left(b-\frac{\la}{2},\la,a \right), & \quad V \sim \mathrm{K}   (\la, a,b  ; 1, \infty )=\mathrm{Ga}\left(b+\frac{\la}{2}, a \right)
\end{align*} 
where $\lambda \in \R$, $a,b >0$, $-b < \frac{\lambda}{2} < b$. \\
(iv) For $F=\tilde{F}^{\epsilon}_{\mathrm{K-Ga},A} $, 
\begin{align*}
& X \sim \mathrm{K} (\la, a, b ; \epsilon,\epsilon^{-1})= \mathrm{K} (\la, a\epsilon, b ; 1,\epsilon^{-1}),  \quad  Y \sim \mathrm{K} (-\la, a, b ; \epsilon, 0)=\mathrm{K} (-\la, a\epsilon, b ; 1, 0) , \\
& U \sim \mathrm{K} (-\la,  a,b ; \epsilon, \epsilon^{-1})=  \mathrm{K} (-\la,  a\epsilon, b ; 1, \epsilon^{-1}),  \quad V \sim \mathrm{K} (\la, a,b  ; 0, 1) =\mathrm{Be}' \left(b+\frac{\la}{2},-\la \right)
\end{align*} 
where $\lambda <0$, $a,b >0$, $-b < \frac{\lambda}{2}$.
By changing parameters $a \to \epsilon^{-1}a$ and taking $\epsilon=0$, for $F=F^+_{\mathrm{Ga}}$,
\begin{align*}
X \sim \mathrm{K}  (\la, a, b ; 1, \infty) =\mathrm{Ga} \left(b+\frac{\la}{2}, a \right) , & \quad  Y \sim \mathrm{K}  (-\la, a, b ; 1,0)=\mathrm{Ga}(-\la,a) \\
U \sim \mathrm{K}   (-\la, a,b ; 1, \infty)= \mathrm{Ga}\left(b-\frac{\la}{2}, a \right), & \quad V \sim \mathrm{K}   (\la, a, b  ; 0, 1) =\mathrm{Be}' \left(b+\frac{\la}{2},-\la \right)
\end{align*} 
where $\lambda <0$, $a,b >0$, $-b < \frac{\lambda}{2}$.\\
(v) For $F=F^{+,\a,\b}_{\mathrm{GIG}}$,
\begin{align*}
X \sim \mathrm{GIG} (\la, a, b ; \a, 1)=\mathrm{GIG}(\la, a \a, b) , & \quad  Y \sim \mathrm{GIG} (\la, b,a ; \b, 1) =  \mathrm{GIG}(\la, b \b,a), \\
U \sim \mathrm{GIG} (\la,  b,a ; \a, 1)=\mathrm{GIG}(\la, b \a, b), & \quad V \sim \mathrm{GIG} (\la,a,b  ; \b, 1)= \mathrm{GIG}(\la, a \b, b)
\end{align*} 
where $\lambda \in \R$, $a,b >0$.
\end{cor}

This recovers all known results on the IP property except for $F_{\mathrm{N}}$ and $F_{\mathrm{C}}$ and the zero-temperature versions. Though we do not discuss here, the zero-temperature limits of the above subtraction-free bijections and their (continuous) solutions are also systematically obtained. Hence, combining with Lemma \ref{lem:scale}, we can conclude that the most fundamental IP property is for $H_I^{+}$ and all other results except $F_{\mathrm{N}}$ and $F_{\mathrm{C}}$ are derived by changing variables, taking special parameters and performing some limiting procedures from that for $H_I^{+}$. This explains why the solutions have exactly three parameters for all bijections in Corollary \ref{cor:ip}, which was a mystery until now. 

\begin{rem}
In Corollary \ref{cor:ip}, we did not apply the relation (v) of Theorem \ref{thm:relation} since the distribution $\mathrm{K}(\la,a,b ; p,q)$ are not closed for the bijection $I(x)=x^{-1}$. If we introduce the inverse Kummer distribution properly, we can also derive the IP property for $F^+_{\mathrm{Ga}}$ using the relation (v) of Theorem \ref{thm:relation}. 
\end{rem}

\section*{Acknowledgements}
This research was supported by JSPS Grant-in-Aid for Scientific Research (B), 19H01792. The authors thank Willox Ralph for helpful comments.

\bibliographystyle{amsplain}
\bibliography{Yang-Baxter}

\end{document}